\documentclass[preprint,showpacs,preprintnumbers,amsmath,amssymb]{revtex4}

\usepackage{graphicx}
\usepackage{dcolumn}
\usepackage{bm}

\begin{document}
\title{Pure, $Si$ and $sp^3$-doped Graphene nanoflakes: a numerical study of density of states}

\author{   N.Olivi-Tran }
\affiliation{  G.E.S., Universite Montpellier II, UMR CNRS 5650
Case courrier 074, place Eugene Bataillon, 34095 Montpellier cedex 05, France}

\date{\today}

\begin{abstract}
We built graphene nanoflakes doped or not with $C$ atoms in the $sp^3$ hybridization or  with $Si$ atoms. These nanoflakes are isolated,
i.e. are not connected to any object (substrate or junction). We used a modified tight binding
method to compute the $\pi$ and $\sigma$ density of states. The nanoflakes are semiconducting (due
to the armchair geometry of their boundaries)
when their are pure but the become conducting when doped because doping removes the degeneracy
of the density of states levels. Moreover, we showed that the $\pi$ Fermi
level and the Fermi level of both $\pi$ and $\sigma$ electrons are not superimposed
for small isolated nanoflakes.

\end{abstract}

\maketitle

\pagebreak
\section{Introduction}
The electronic structure of nanoscale graphene has been the subject of intensive research during
the last years because of fundamental scientific interest in nanomaterials. With the developments
in preparation and synthesis techniques carbon-based nanostructures have emerged as one of the most 
promising materials for non silicon electronics.

Graphene nanoflakes are either semiconductors or conductors aiming to the geometry 
of their boundaries \cite{heiskanen}.
Several studies have been published on graphene nanoflakes
doped with impurities \cite{Pershoguba,Fogler,Pellegrino}. We decided here to compute the $\pi$ and $\sigma$
electrons density of states of graphene nanoflakes doped with a small
percentage of either $Si$ atoms or with $sp^3$ hybridized $C$ atoms.

We show here numerical results on the conductivity of graphene nanoflakes
and on the density of states (mainly the Fermi level) of impurities-doped
graphene nanoflakes.

Our numerical method allows us to take into account the $\pi$ as well as the $\sigma$
for the calculation of the electronic density of states. And moreover, this method allows one
to separate the $\pi$ and the $\sigma$ density of states.

\section{Tight-binding model with $sp^{\nu}$ hybridization}
Let us remind that this is a one electron model, each electron moves in a potential $V(r)$
which represents both the nuclei attraction and the repulsion of the other electrons.
$\sigma$ and $\pi$ electrons are separately treated.

The Hamiltonian of the $\sigma$ bonds may be written as:
\begin{equation}
H_{\sigma}=E_m\sum_{i,J}|iJ><iJ|+\Delta_{\sigma}\sum_{i,J,J' \neq J}|iJ><iJ'|+\beta_{\sigma}
\sum_{i,i'\neq i,J}|iJ><|i'J|
\end{equation}
where the molecular orbital $\sigma$ is given by:
\begin{equation}
|\psi>=\sum_{i,J}a_{i,j}|iJ>
\end{equation}
with $|iJ>$ the hybrid $sp^{\nu}$ orbital ($\nu=1,2,3$) which points from site $i$
along the bond $J$; the energy origin is taken at the vacuum level.
$i$ and $i'$ are first neighbours with $E_m$ the average energy: $E_m=(E_s+\nu E_p)/(\nu+1)$
$E_s$ and $E_p$ are the atomic energy levels; $\beta_{\sigma}$ is the usual hopping
or resonance integral in the tight binding theory (interaction between nearest neighbour atoms
along the bond); $\Delta_{\sigma}$ is a promotion integral (transfer between hybrid orbitals on the same site): $\Delta_{\sigma}=(E_s-E_p)/(\nu+1)$.

For an infinite three dimensional crystal (bulk), the gap between valence and conduction band (forbidden band)
is $g=|-2\beta_{\sigma}+(\nu+1)\Delta_{\sigma}|$ (for VIb elements) and $\beta_\sigma$
may be derived from the values of $g$ \cite{leleyter}.

The Hamiltonian for the $\pi$ bonds is:
\begin{equation}
H_{\pi}=E_p\sum_i|i><i|+\beta_{\pi}\sum_{i, i' \neq i} |i><i'|
\end{equation}
with $|i>$ the $\pi$ orbitals centered on atoms $i$ and $\beta_{\pi}$ is the hopping integral
for $\pi$ levels.

The $\beta_{\pi}$ value for $C$ was chosen in order to get the correct positions for $C_2$ energy levels in comparison to the results of the Verhaegen's ab initio calculations \cite{verhaegen}.

We need only three parameters:$\beta_{\sigma}$, $\beta_{\pi}$ and $\Delta_{\sigma}$ for the
homonuclear model which represent in fact the average potential $V(r)$ and which take into
account the nuclear attraction and the dielectronic attraction \cite{leleyter}.
The values of these three parameters are given in table I.

\section{Results}
We computed the density of states of pure and doped graphene nanoflakes.
The nanoflakes have boundaries represented in figure 1.
We studied nanoflakes containing 480, 720 and 960 atoms: we obtained their geometry
by reproducing figure 1 in the $y$ direction.

In figure 
2, one may see the density of states of nanoflakes of graphene containing
480 atoms. The nanoflakes are isolated, i.e. they are not connected to any other
conducting material or connected to any substrate.
The 3 graphs in figure 2 representing the DOS (density of states) correspond
to: (a) pure graphene nanoflake (b) 5\% of Si atoms randomly introduced in the graphene nanoflake and
(c) 5\% of $sp^3$ hybridized carbon atoms randomly distributed within the nanoflake.
, 10\%, 15\% and 20\%.
The boundaries of the nanoflakes correspond to the armchair geometry.

Figure 3 corresponds to the same nanoflakes (i.e. pure, with 5\% of $Si$ atoms and with 5\%
of $sp^3$ hybridized carbon atoms) but in this case for 720 atoms.

Figure 4 corresponds also to 3 DOS with the same nanoflakes but this time containing 960 atoms.

It is possible to introduce impurities like $Si$ atoms. Carbon and Silicon belong to the same column IVb
of the periodic table. However, their structures in the solid states are different.
Carbon indeed can be found in several allotropic varieties such as diamond, graphite, graphene
whereas silicon is always found in three dimensional systems.

In mixed silicon-carbon $Si_nC_p$ nanoclusters, it has been previously pointed out
that experimental results show a progressive transition from a carbon type behaviour
of Si to a silicon type behaviour \cite{dornenburg} as the silicon richness is growing in the clusters \cite{leleyter6}.
This may be transferred in nanoflakes to the percentage of $Si$: here we deal with small percentages
of $Si$ surrounded by  95\% of $C$ at least. From references \cite{dornenburg,leleyter6} we may say that due to the richness
in carbon atoms in the studied nanoflakes of graphene, the $Si$ atoms adopt the carbon behaviour
i.e. are in the $sp^2$ hybridization.

\section{Discussion}
It is worth noticing that for small and finite size nanoflakes, the DOS consists
of many discrete levels and instead of Fermi level, one should speak of highest occupied
molecular orbital (HOMO) and lowest unoccupied molecular orbital (LUMO).

However, we retain these terms in a loose sense because the main result when comparing the three figures of DOS for the nanoflakes of graphene
of different sizes,  
we observe (figures 2,3 and 4) that the Fermi (or HOMO) level of the $\pi$ electrons DOS
correspond to the DOS in the $k=0$ direction in the Brillouin zone (see for that reference 
\cite{wakabayashi}. The fact is that for 480 atoms and 720 atoms of pure graphene (figures 2(a) and
3(a)) the DOS shows that there is a gap between the $\pi$ Fermi level and higher $\pi$ energy levels:
this indicates that the nanoflake of pure $C$ atoms is semiconducting. This is in good
agreement with the literature showing that with armchair boundaries (which is the case
in our nanoflakes) semiconducting is generally obtained \cite{heiskanen}.

By doping the nanoflake with either $Si$ atoms or by $sp^3$ carbon atoms, one destroys the
semiconducting character. This is due to the fact that impurities remove the degeneracy
of the energy levels.

What is also interesting is that the $\pi$ 
Fermi level and the whole Fermi level are not superimposed for nanoflakes containing
a small number of atoms (480 and 720 atoms: fig. 2 and 3). The superimposition only occurs for a larger
number of atoms (here 960 atoms: fig.4).
Generally in the literature, one studies the $\pi$ DOS as the major numerical methods
used by authors is the tight binding one taking only account of $\pi$ electrons.
But here, we deal with $\sigma$ and $\pi$ atoms, and when calculating the Fermi the $\pi$
and the whole Fermi level, these two do not superimpose, at least for a small number
of atoms. But  $\sigma$ electrons do not enter in the conducting or semiconducting
character of the sample.

\section{Conclusion}
Let us summarize the results: doping graphene nanoflakes with armchair boundaries
removes the degeneracy of the energy levels in the DOS and therefore removes
the semiconducting character of graphene nanoflakes with this kind of  boundaries; the $\pi$ Fermi level and joint $\pi$ and $\sigma$ Fermi level do not superimpose for small graphene nanoflakes: but as the $\sigma$ electrons
do not enter  conduction this has no consequence on the electronic features of small 
graphene nanoflakes.
\pagebreak

\pagebreak
\begin{table}
\begin{center}
\begin{tabular}{|c|c|c|c|c|}
\hline
 atom/parameter & $E_{\sigma}$ & $E_{\pi}$ & $\beta_{\sigma}$ & $\beta_{\pi}$ \\
\hline
$C$ & 19.45 & 10.74 & 7.03 & 3.07 \\
\hline
$Si$ & 14.96 & 7.75 & 4.17 & 0.8 \\
\hline 
\end{tabular}
\end{center}
\caption{Parameters for the tight binding calculations in $eV$}
\end{table}

\begin{figure}
\includegraphics[width=10cm]{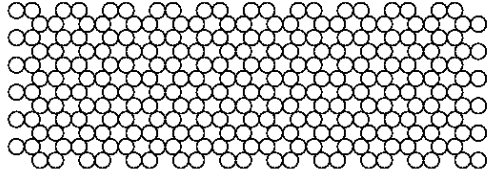}
\caption{Example of graphene nanoflake with 240 $C$ atoms. Our graphene nanoflakes are obtained
by adding the same structure in the $y$ direction}
\end{figure}
\begin{figure}
\includegraphics[width=10cm]{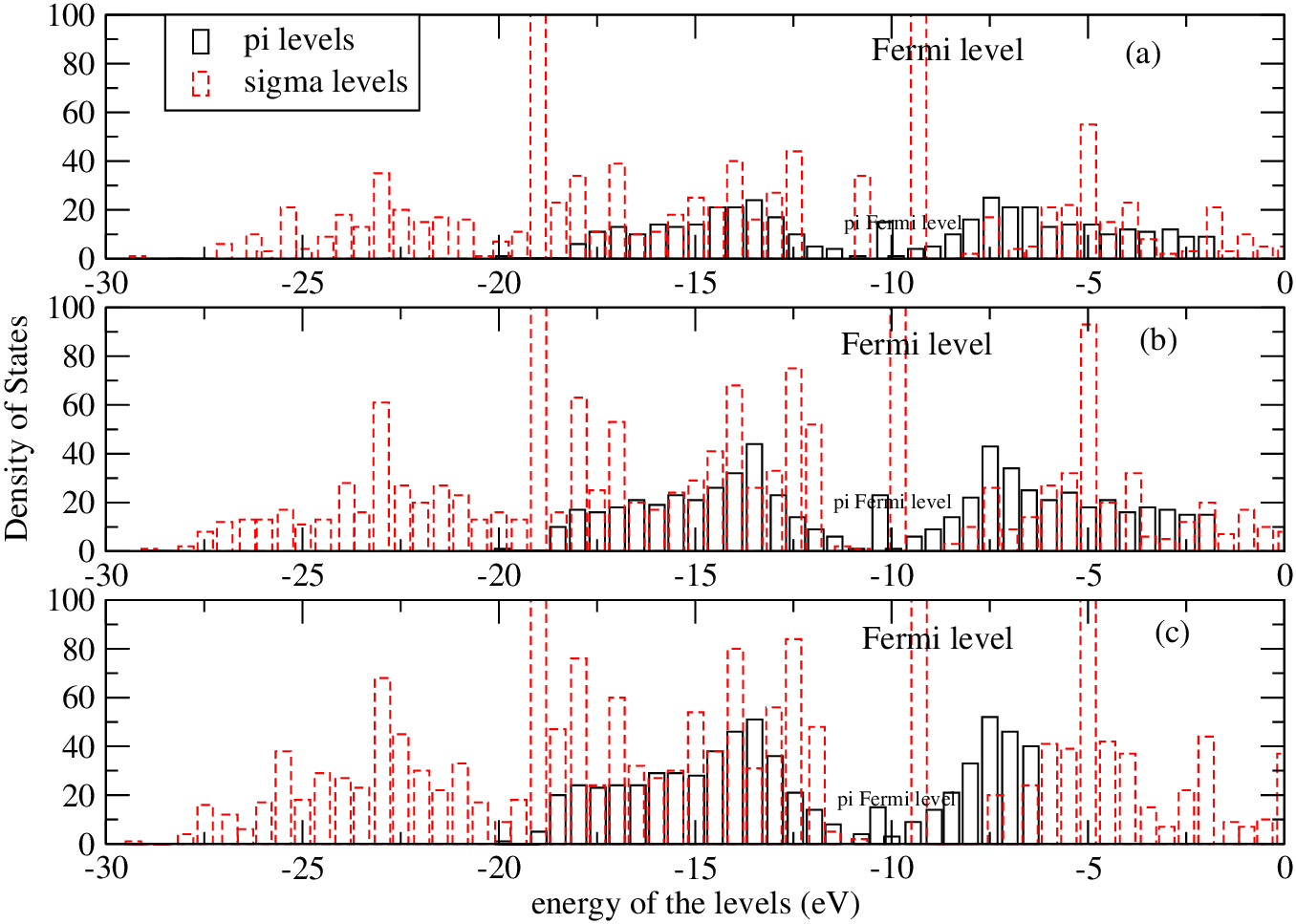}
\caption{Electronic density of states of a graphene nanoflake containing 480 atoms. (a) pure $c$; (b)
doped with 5\% of $Si$ atoms; (c) doped with \%5 of $C$ atoms in the $sp^3$ hybridization}
\end{figure}
\begin{figure}
\includegraphics[width=10cm]{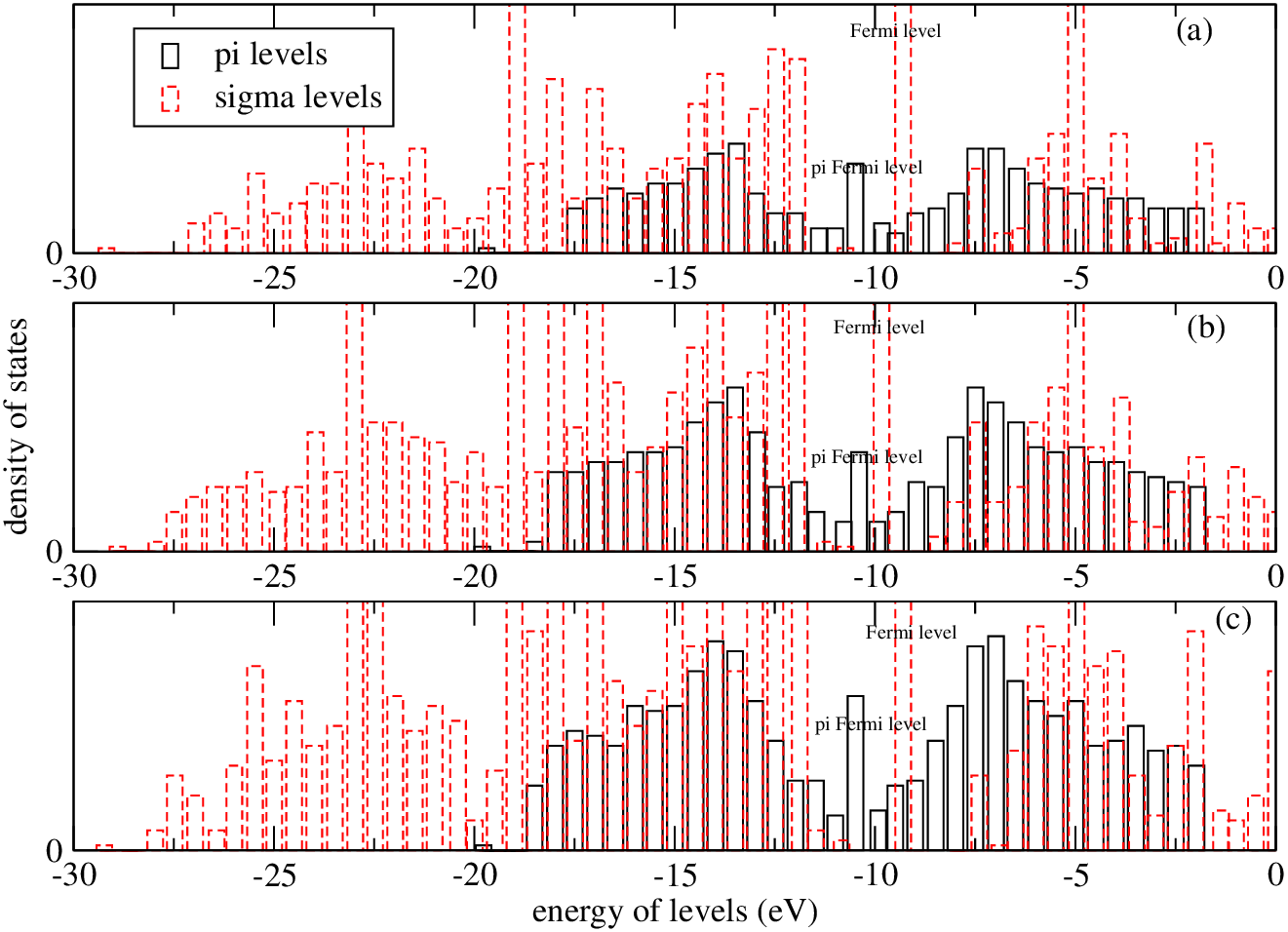}
\caption{Electronic density of states of a graphene nanoflake containing 720 atoms. (a) pure $c$; (b)
doped with 5\% of $Si$ atoms; (c) doped with \%5 of $C$ atoms in the $sp^3$ hybridization
}
\end{figure}
\begin{figure}
\includegraphics[width=10cm]{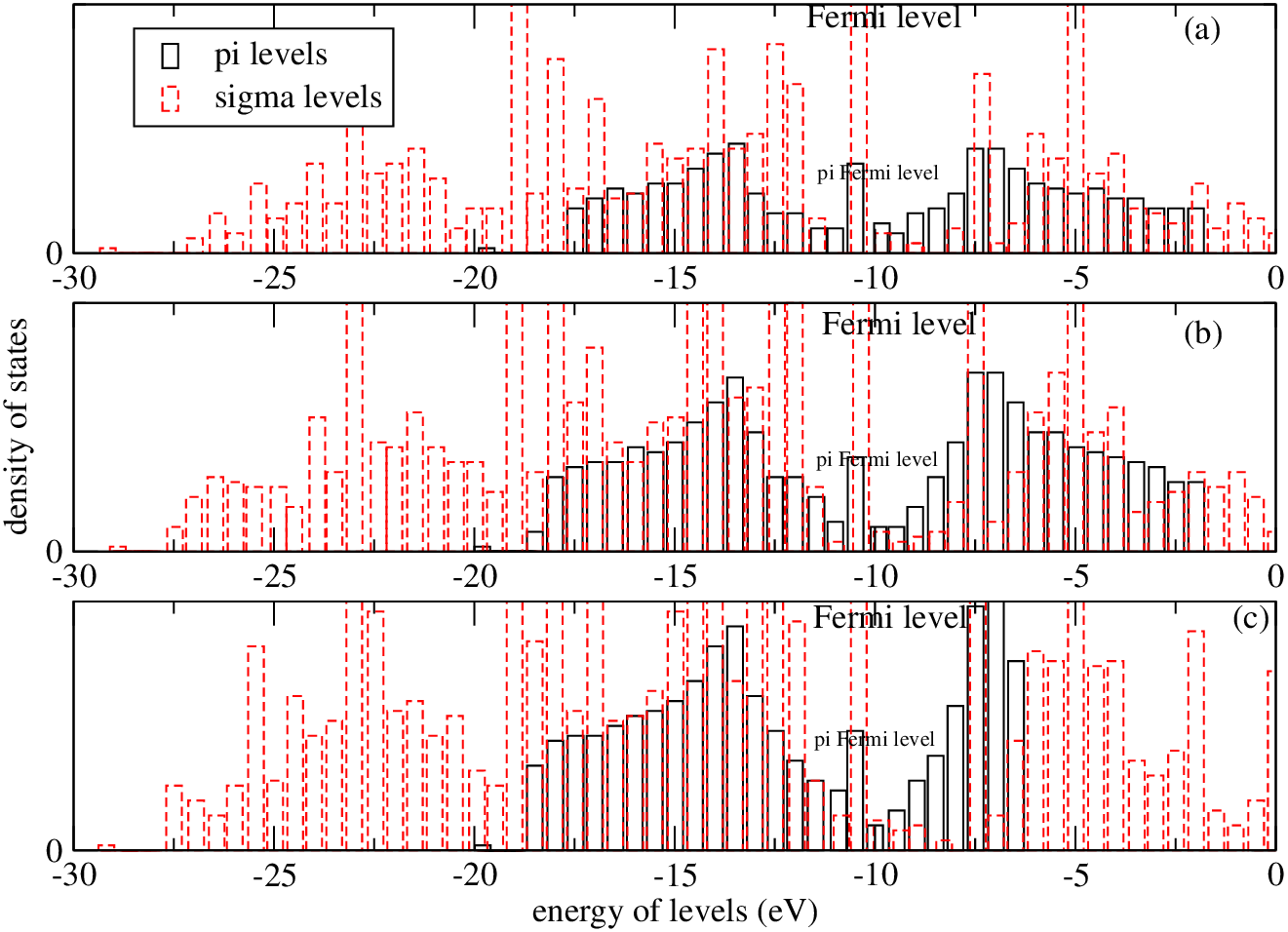}
\caption{Electronic density of states of a graphene nanoflake containing 960 atoms. (a) pure $c$; (b)
doped with 5\% of $Si$ atoms; (c) doped with \%5 of $C$ atoms in the $sp^3$ hybridization}
\end{figure}

\end{document}